\documentclass[twocolumn]{aastex62}

\graphicspath{{./}{figures/}}

\submitjournal{ApJL}

\shorttitle{Symmetry broken in AGN disks}
\shortauthors{Wang et al.}

\begin{document}

\title{SYMMETRY BREAKING IN DYNAMICAL ENCOUNTERS IN THE DISKS OF ACTIVE GALACTIC NUCLEI}



\correspondingauthor{Yi-Han Wang}
\email{yihan.wang.1@stonybrook.edu}

\author[0000-0002-8614-8721]{Yi-Han Wang} 
\affiliation{Department of Physics and Astronomy, Stony Brook University, Stony Brook, NY 11794-3800, USA}

\author[0000-0002-9726-0508]{Barry McKernan}
\affiliation{Department of Science, BMCC, City University of New York, New York, NY 10007, USA}
\affiliation{Center for Computational Astrophysics, Flatiron Institute, New York, NY 10010, USA}

\author{Saavik Ford}
\affiliation{Department of Science, BMCC, City University of New York, New York, NY 10007, USA}
\affiliation{Center for Computational Astrophysics, Flatiron Institute, New York, NY 10010, USA}

\author{Rosalba Perna}
\affiliation{Department of Physics and Astronomy, Stony Brook University, Stony Brook, NY 11794-3800, USA}
\affiliation{Center for Computational Astrophysics, Flatiron Institute, New York, NY 10010, USA}

\author{Nathan W. C. Leigh}
\affiliation{Department of Physics and Astronomy, Stony Brook University, Stony Brook, NY 11794-3800, USA}
\affiliation{Department of Astrophysics, American Museum of Natural History, New York, NY 10024, USA}
\affiliation{Departamento de Astronom\'ia, Facultad Ciencias F\'isicas y Matem\'aticas, Universidad de Concepci\'on, Av. Esteban Iturra s/n Barrio Universitario, Casilla 160-C, Concepci\'on, Chile}

\author[0000-0003-0064-4060]{Mordecai-Mark Mac Low}
\affiliation{Department of Astrophysics, American Museum of Natural History, New York, NY 10024, USA}



\begin{abstract}
Active galactic nucleus (AGN) disks may be important sites of binary black hole (BBH) mergers. Here we show via numerical experiments with the high-accuracy, high precision code {\tt SpaceHub} that broken symmetry in dynamical encounters in AGN disks can lead to an asymmetry between prograde and retrograde BBH mergers. The direction of the hardening asymmetry depends on the initial binary semi-major axis. An asymmetric distribution of mass-weighted projected spin $\chi_{\rm eff}$ should therefore be expected in LIGO-Virgo detections of BBH mergers from AGN disks. This channel further predicts that negative $\chi_{\rm eff}$ BBH mergers are most likely for massive binaries.

\end{abstract}

\keywords{LIGO; Active galactic nuclei: Black hole physics; Stellar dynamics; Stellar mass black holes; Supermassive black holes; Gravitational waves}

\section{Introduction} 
Active galactic nucleus (AGN) disks may be important sites for stellar mass binary black hole (BBH) mergers \citep{McK12,Bartos17,Stone17,Leigh18,Samsing20}, but the detailed processes that lead to a BBH merger in an AGN disk are not yet well-constrained \citep[e.g.][]{McK18,Grobner20}. We expect that binary formation in AGN disks is extremely efficient \citep{Secunda20a,Tagawa20b}, and that initial binary hardening will occur due to gas drag (e.g. \citealt{Baruteau11}, but see also \citealt{Li21,Tiede20}). However, we expect gas drag will eventually become inefficient \citep[e.g.][]{Leigh14}, leading to binaries that stall at semi-major axes 
    too large for a gravitational wave-driven merger to occur
in less than the AGN disk lifetime. Thus, dynamical encounters could play a critical role in hardening (or disrupting) binaries in the AGN channel \citep[e.g.][]{Leigh18,Secunda20a,Samsing20,Tagawa20c}. The imprint of dynamical processes particular to AGN disks could appear in the gravitational wave signal of such BBH mergers.

\begin{figure}
\begin{center}
\includegraphics[width=0.95\linewidth,angle=0]{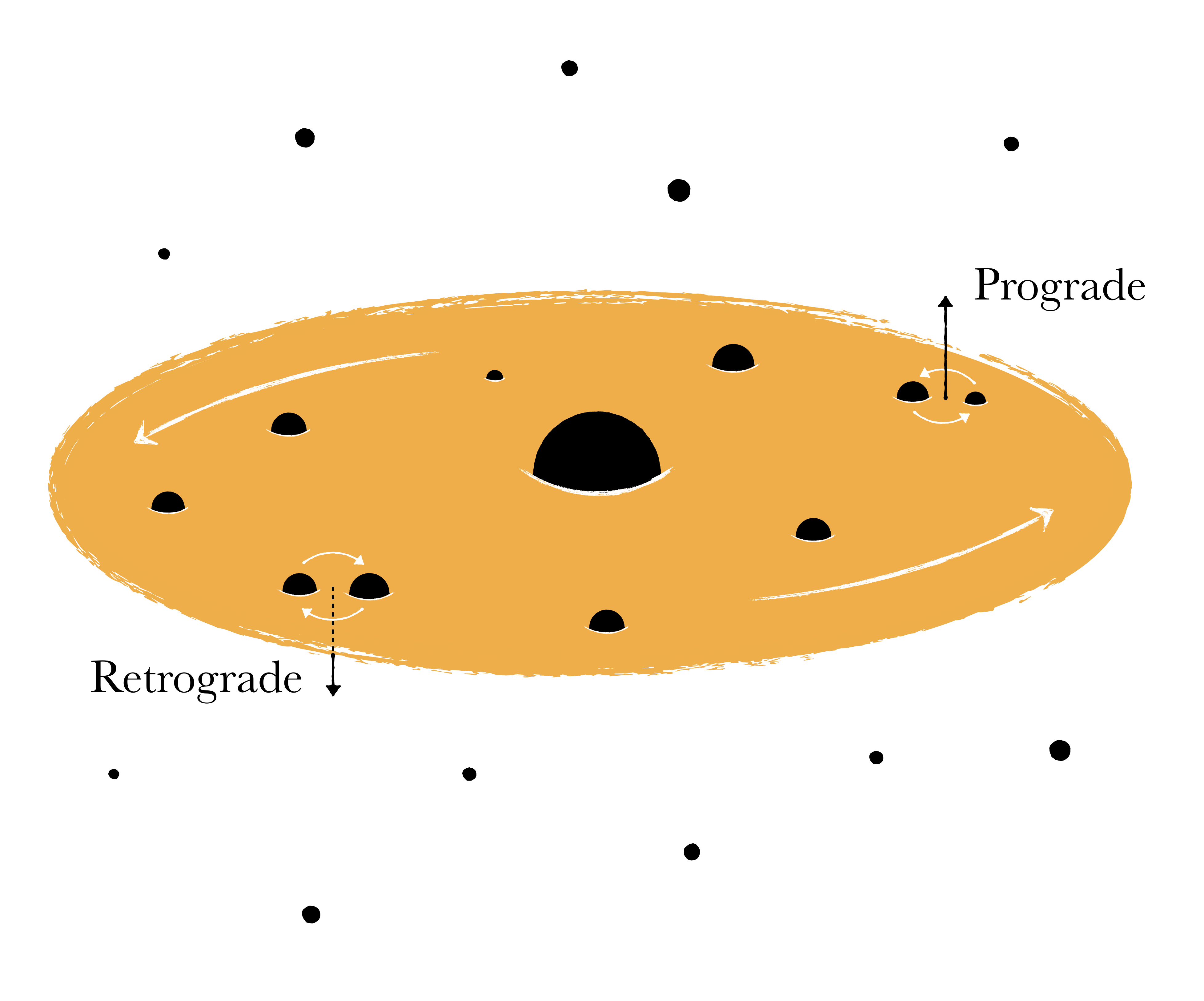}
\end{center}
\caption{Prograde versus Retrograde Binaries: Single BHs in an AGN disk that are orbiting the SMBH with their angular momentum aligned to the disk angular momentum $\vec{L}_{\rm d}$ will form binaries whose center of mass will continue to orbit with the disk gas. However, the orbital angular momentum of the binary around its own center of mass may be either prograde (aligned with $\vec{L}_{\rm d}$) \textit{or} retrograde (anti-aligned with $\vec{L}_{\rm d}$). We use the terms `prograde' ($+$) and `retrograde' ($-$) binary, respectively, for these two arrangements. (Figure credit: T. Callister, personal comm.).} 
\label{fig:retrodiagram}
\end{figure}

Within the LIGO-Virgo detected population, there is an intriguing possible anti-correlation between the mass ratio
    of binaries with masses $M_1$ and $M_2$ 
and the mass-weighted projection of spin onto binary orbital angular momentum ($\chi_{\rm eff}$) in BBH mergers \citep{Tom2021}. Most dynamical channels for BBH mergers are rotationally symmetric and therefore predict a symmetric distribution of $\chi_{\rm eff}$ around $\chi_{\rm eff}=0$ \citep[e.g.][]{Carl2018, Liu2017}. However, \citet{qXeff} point out that there are several natural sources of symmetry breaking in the AGN channel for BBH mergers. Figure~\ref{fig:retrodiagram} shows BBHs in an AGN disk with binary angular momenta oriented (anti-)parallel to the disk angular momentum and so  (retrograde) prograde compared to the disk gas angular momentum. Hereafter we denote prograde and retrograde binaries as ($+$) or ($-$) respectively and prograde/retrograde tertiaries as $+/-$.

\citet{qXeff} 
     argued
 that: 1) the hard-soft boundary for ($+$) and ($-$) binaries should be different for tertiary encounters from a preferred direction (i.e. different hard-soft boundaries for ($+$),$+$ and ($-$),$+$ and 2) more massive ($-$) binaries are more probable survivors of $+$ tertiary encounters. In this Letter we test and quantify these hypotheses via numerical scattering experiments.

\section{Numerical methods}

The scattering experiments are performed  using our high-precision, few-body code {\tt SpaceHub} \citep{Wang2021SpaceHub}, which employs several novel algorithms shown to  out-perform other numerical methods in the literature for both accuracy and precision.   
For the scattering experiments performed for this work, since the mass ratio between the stellar mass black hole (BH) and the supermassive black hole (SMBH) is very small, and the scattering may involve extreme eccentricities, we use the AR-chain$^+$ method in {\tt SpaceHub}, which can deal with systems with extreme eccentricity and low mass ratios accurately and precisely. Post-Newtonian general relativity terms are included.  We include the first order (precession), second order (correction to precession), and 2.5th order terms in the acceleration (gravitational radiation).

We setup coplanar scattering experiments between equal mass prograde ($+$) or retrograde ($-$) BBHs, 
where each BH  in the binary has a mass $30~M_\odot$,
and a single $+$ tertiary BH with mass 10, 30, or 60 M$_\odot$. The relative velocity $v_\infty$ between the single BH and the BBH is assumed to be 50~km/s (the order of magnitude expected for encounters in an AGN disk). A 10$^8$~M$_\odot$ central SMBH is included.
We explored  different distances between the SMBH and the BBH of 16, 8, 4, and 2 R$_{\rm TDE}$, where
\begin{equation}
    R_{\rm TDE} = \bigg(\frac{M_{\rm SMBH}}{M_{\rm BBH}}\bigg)^{1/3}a_{\rm BBH}
\end{equation}
is the tidal disruption radius of the BBH, which can be parameterized as
\begin{equation}
    R_{\rm TDE} \sim 120r_{g}\left(\frac{M_{\rm SMBH}}{10^{8}M_{\odot}}\right)^{1/3} \left(\frac{M_{\rm BBH}}{60M_{\odot}}\right)^{-1/3}\left(\frac{a_{\rm BBH}}{1{\rm AU}}\right)
\end{equation}
for $r_{g}=GM_{\rm SMBH}/c^{2} \sim 1\mbox{ AU}(M_{\rm SMBH}/10^{8}M_{\odot})$. Thus 2--16$R_{\rm TDE}$ spans $\sim 200$--$2000r_{g}$ in a disk around an $M_{\rm SMBH} \sim 10^{8}M_{\odot}$ SMBH. This includes regions where a migration trap may exist in the disk (\citealt{Bellovary16,Samsing20,Secunda20a,Yang19,Tagawa20}, but see also \citealt{Dittmann2020}). The range of distances between the SMBH and the binary is chosen so that it straddles the two limits of the binary being tidally disrupted by the SMBH due to the Hills mechanism, and the SMBH not having any significant influence on the scattering event.  
The latter limit can be seen by comparison with
reference scattering experiments which we
further performed  without the SMBH.

\section{Results from scattering experiments}\label{sec:results}

\begin{figure*}
\includegraphics[width=0.5\linewidth]{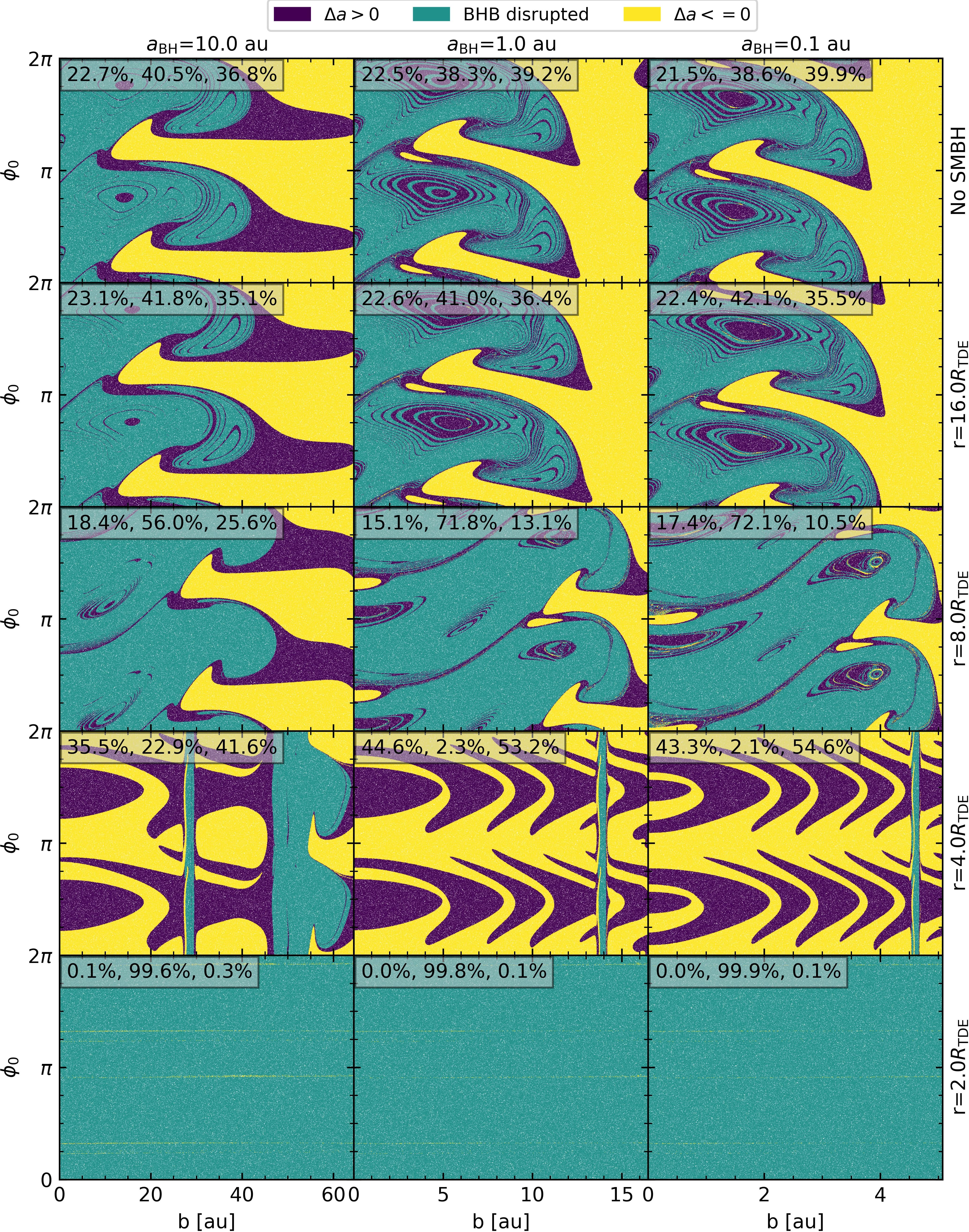}
\includegraphics[width=0.5\linewidth]{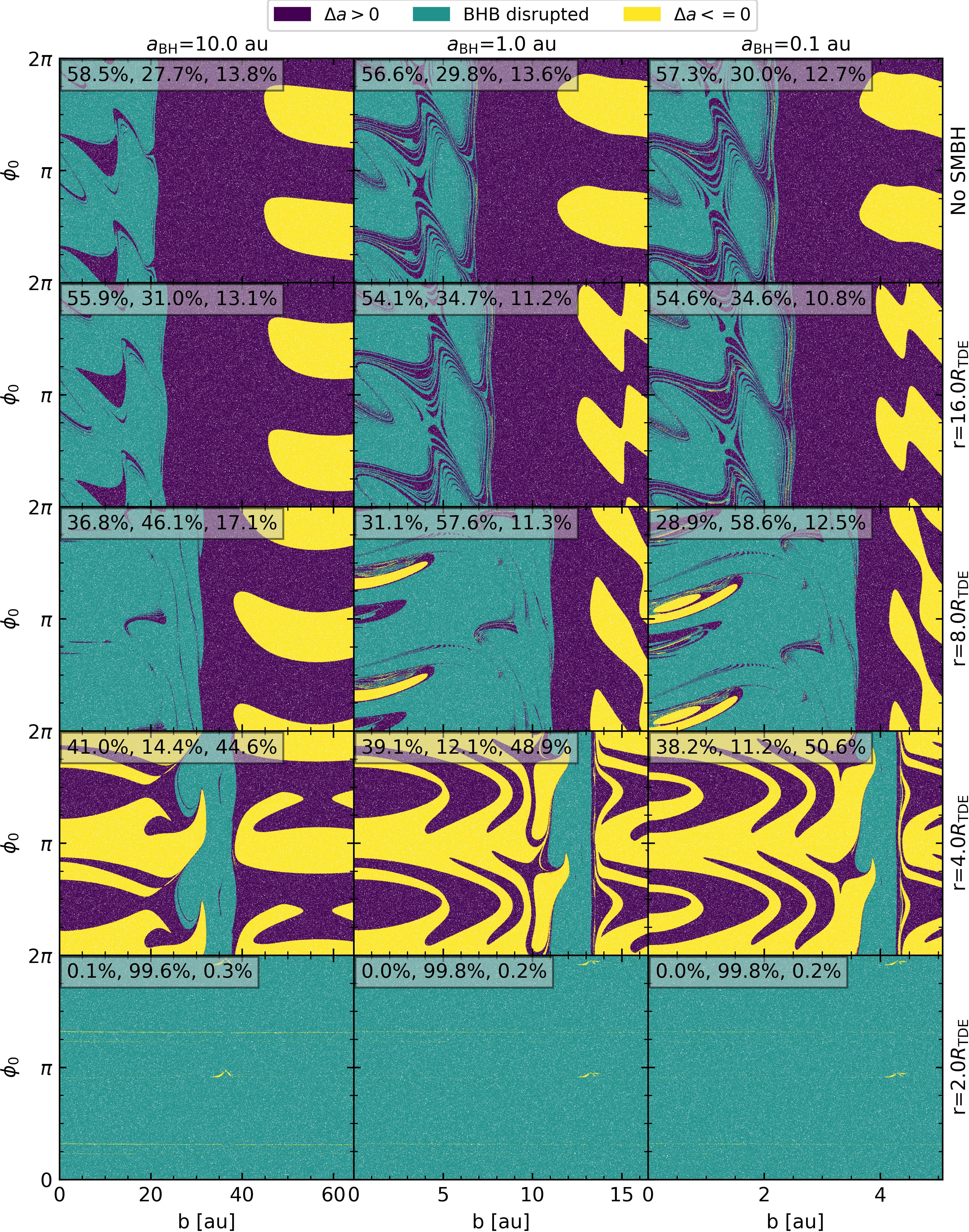}
\caption{Outcome of the scatterings  between a ($+$ LHS; $-$ RHS) BBH and a $+$ tertiary BH, as a function of the phase of the BBH $\phi_0$ and impact parameter $b$ of the tertiary BH. The masses of the BHs in the ($+,-$) binary are both $30 M_\odot$, while the mass of the $+$ tertiary is also 30$M_\odot$. The {\rm left} panels show the results for a $(+),+$ encounter, while the {\rm right} panels show the results for a $(-),+$ encounter. 
    The radius of the encounter from the central SMBH is shown on the right.  The percentage of encounters that soften, disrupt, or harden the BBH are given at the top of each subplot.
    }
\label{fig:M30}
\end{figure*}

\begin{figure*}
\includegraphics[width=0.5\linewidth]{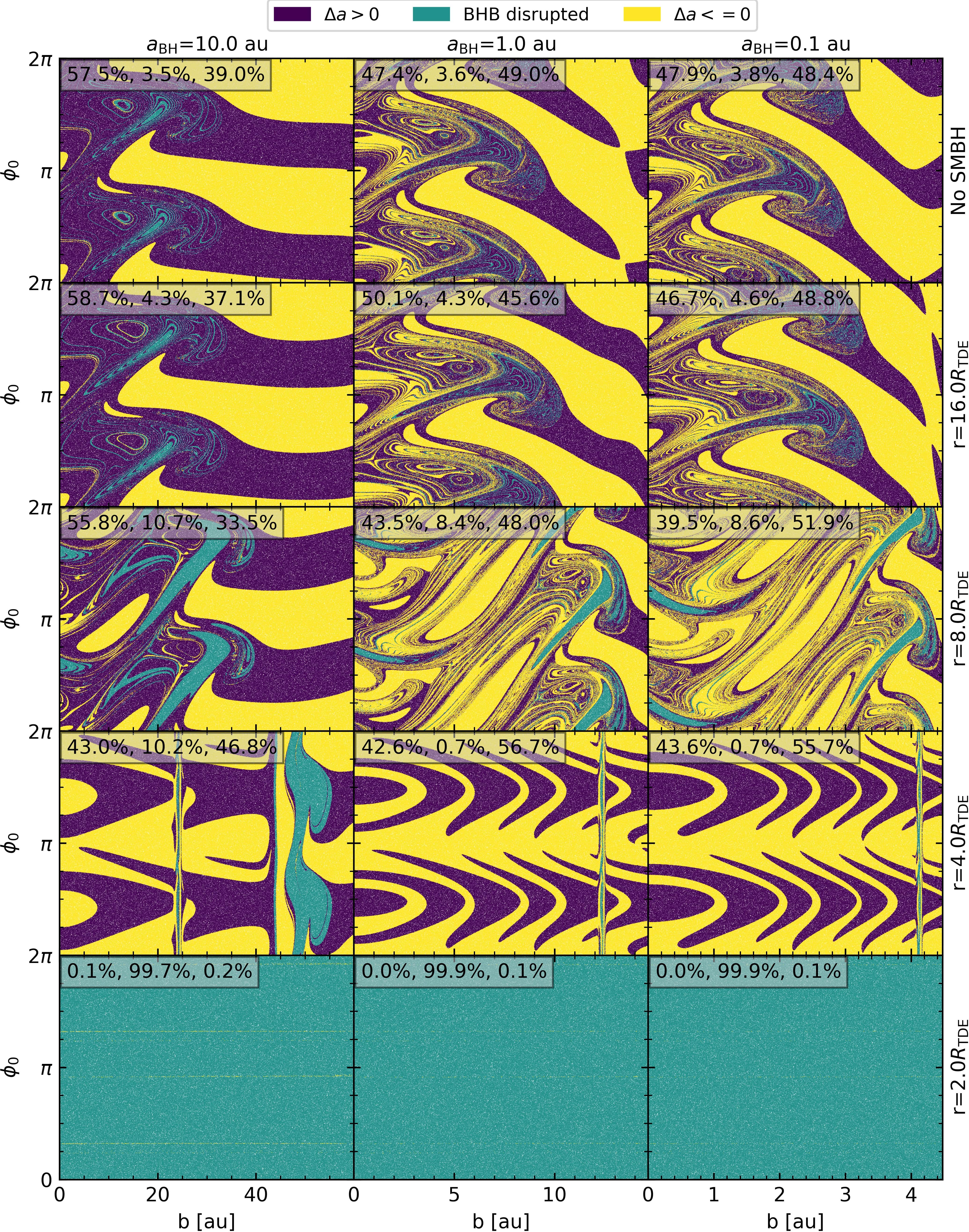}
\includegraphics[width=0.5\linewidth]{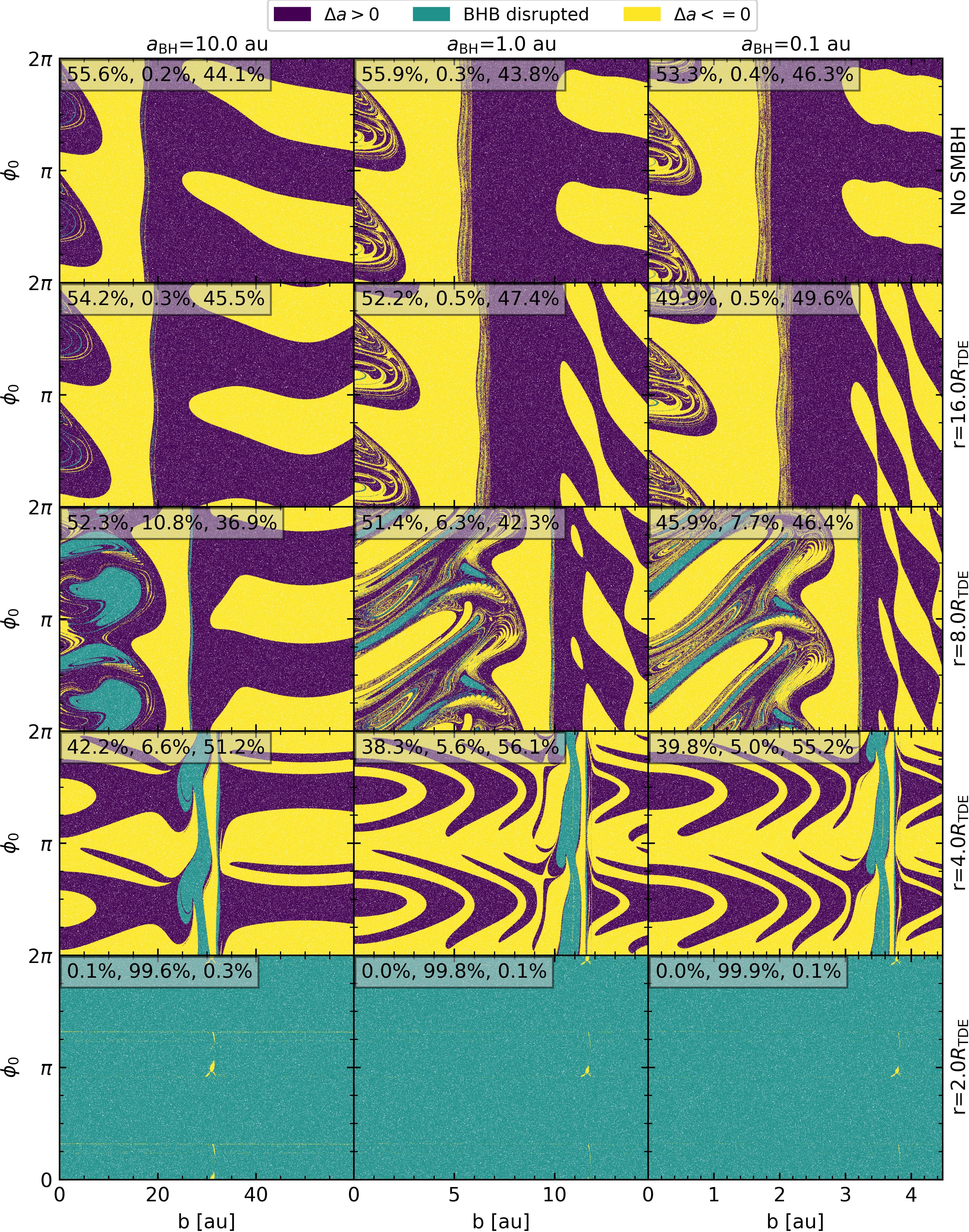}
\caption{Same as in Fig.~\ref{fig:M30} but for a lower mass $+$ tertiary with mass 10~$M_\odot$.}
\label{fig:M10}
\end{figure*}

\begin{figure*}
\includegraphics[width=0.5\linewidth]{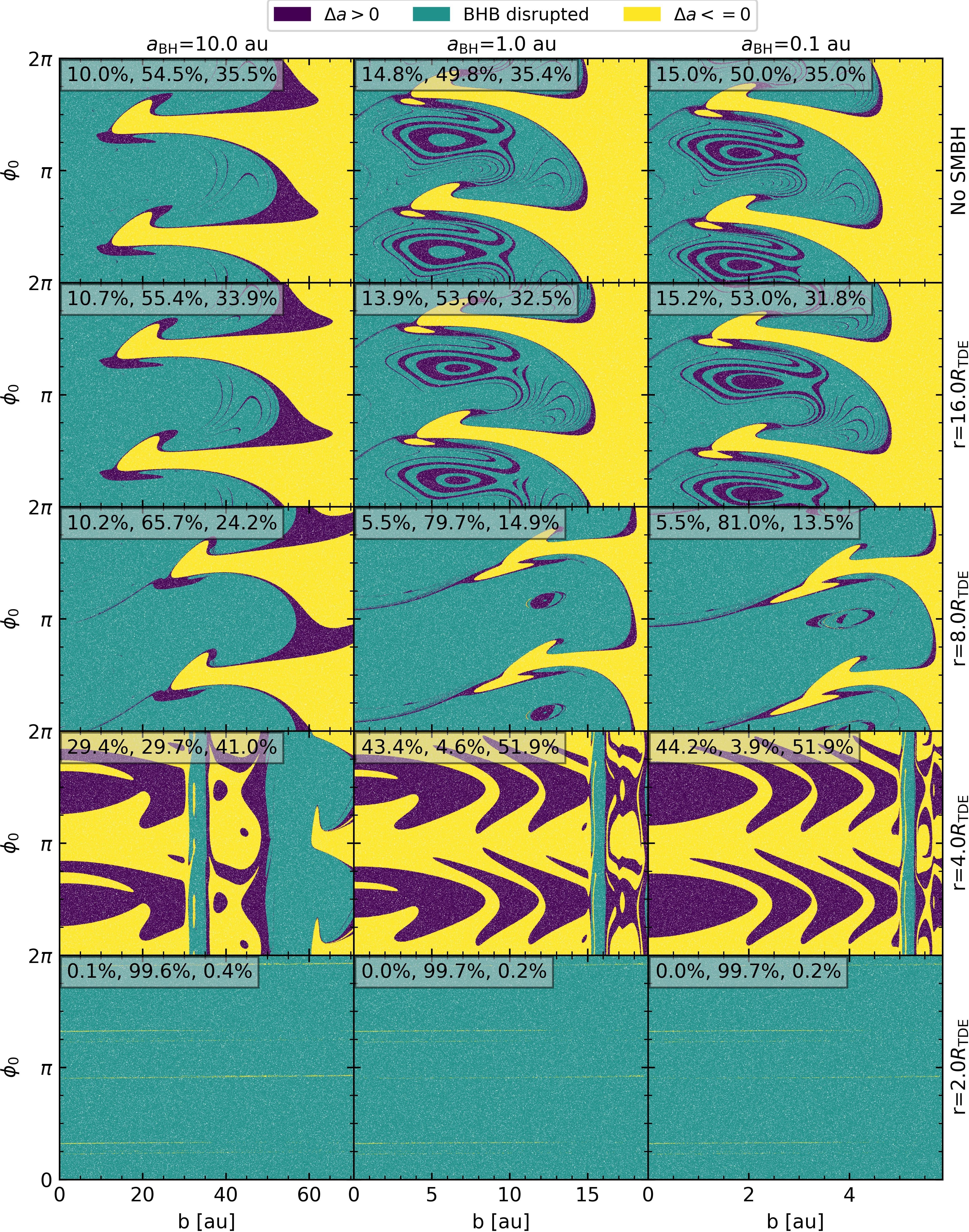}
\includegraphics[width=0.5\linewidth]{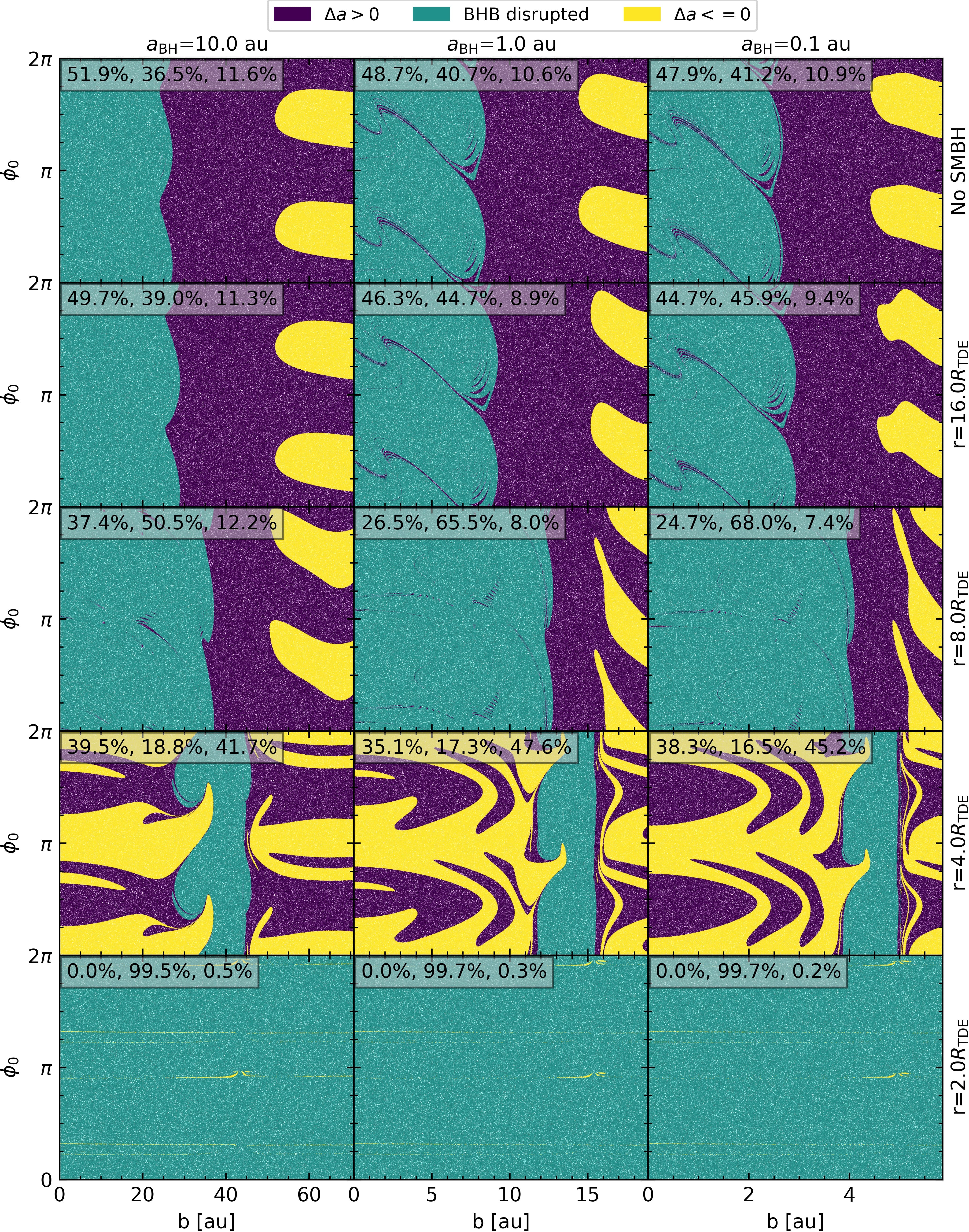}
\caption{Same as in Fig.~\ref{fig:M10} but for a higher mass tertiary with mass 60~$M_\odot$.}
\label{fig:M60}
\end{figure*}

The results of our scattering experiments are displayed in 
Figures~\ref{fig:M30}, \ref{fig:M10}, and \ref{fig:M60} for the three values of the mass of the tertiary BH: 30, 10, and 60~$M_\odot$, respectively. 

In the case of an equal mass binary and $+$ tertiary in Figure~\ref{fig:M30},  there is a clear visual difference between prograde ($+$) binaries on the left-hand side (LHS) and retrograde ($-$) binaries on the right-hand side (RHS). Consistently, there is a larger region of binary hardening parameter space (in yellow) among $+$ tertiary encounters with ($+$) binaries (LHS), compared with the same region for ($-$) binaries (RHS). The difference corresponds to a roughly $3:1$ hardening ratio across phase space for most of the encounter conditions considered here. We interpret this as due to a different hard-soft boundary for ($+$) and ($-$) binaries relative to $+$ encounters. On average, a $+$ tertiary encounter with a ($-$) binary contains more relative kinetic energy at encounter due to the larger velocity differential. 
In particular, close $+$ encounters (at relatively small impact parameters) are far more likely to soften ($-$) binaries than ($+$) binaries.

We see a difference in outcomes when we drop the mass of the $+$ tertiary encounter to 10$M_{\odot}$ in Fig.\ref{fig:M10}. In this case, the hardening fraction of parameter space (again in yellow) is similar for both ($+$) and ($-$) binaries in most cases. This suggests that ($-$) binaries have sufficient inertia that the relative velocity difference makes less of a difference to the range of chaotic hardening encounters. Thus, a ($-$) binary that is significantly more massive than the $+$ tertiary during encounters is much more likely to survive and harden as a result of dynamical interactions. Finally,  encounters with more massive $+$ tertiaries (60$M_{\odot}$) lead to a higher rate of binary softening or ionization among both the ($+$) and ($-$) binaries, but the hardening ratio (yellow) across all of parameter space remains similar at around $\sim 3:1$. 

The trends discussed above are further modified if the scattering happens near the SMBH. At large disk radii (second row from top), the trends are similar to the case of no SMBH: ($+$) binaries are more likely disrupted in the encounters, whereas ($-$) binaries are more likely softened. At small impact parameters, the encounter is \emph{always} ionizing or softening for ($-$) binaries, whereas for ($+$) binaries, small impact parameters can still lead to binary hardening. 
However, we can see that, as the distance of the BBH from the SMBH gets smaller (i.e. moving toward the lower panels of Figs.~\ref{fig:M10}, \ref{fig:M30} and \ref{fig:M60}), the tidal force from the SMBH starts to dominate the BBH system over the scatterings from the $+$ tertiary, leading to more extended softening areas in these subplots (due to tidal softening by the SMBH). As the distance from the SMBH approaches 2 $R_{\rm TDE}$, the BBH fate is dominated by tidal disruption (as expected).

If we define $A_{\rm hard}$ as the area of phase space in which a binary is hardened, then we can compare $A_{\rm hard,(+),+}$, the fraction of ($+$) BBHs that are hardened by a $+$ tertiary encounter, to $A_{\rm hard,(-),+}$, the fraction of ($-$) BBHs that are hardened by a $+$ tertiary encounter.
This ratio is shown in the upper panel of Figure~\ref{fig:ratios}. The bottom panel of the same figure shows the same area ratios, but including only those binaries whose merger times due to gravitational radiation are smaller than a Hubble time. The three panels of the figure separately show the ratio for the three different masses of the $+$ tertiary considered here. 
Note that some points for $a_{\rm BH}=10$~au are missing from the middle and right panels due to the corresponding areas of the $-$  binaries being infinitesimally small
(so the ratio cannot be defined). 

As indicated by the bottom panel of Fig.~\ref{fig:ratios}, for tight BBHs with 0.1~au semi-major axes, the ratio $A_{\rm hard,(+),+}/A_{\rm hard,(-),+}$ is very close to symmetric ($\sim 1$) compared to larger semi-major axes. This is because the outcomes of the scatterings strongly depend on the relative velocity between the tertiary $+$ perturber and the individual stellar BH in the BBH at closest approach. For tight binaries with 0.1~au semi-major axes, the orbital velocity of the stellar BH ($\sim 700$~km/s) is much higher than the velocity of the tertiary perturber ($\sim50$ km/s); therefore, the relative velocity is rather determined by the orbital velocity of the BBH, and thus the contribution from the tertiary with respect to the $+/-$ motion is very small. Therefore, the $+$ and $-$ scatterings show nearly symmetric results, nearly independent of the mass of the $+$ tertiary BH and the distance to the SMBH. 

The symmetry breaks for BBHs with larger semi-major axes, since, as $a_{\rm BH}$ increases, the relative velocity becomes increasingly dominated by the velocity of the tertiary. The orbital velocity of the 10~au semi-major axis BBH is $\sim 70 {\rm km/s}$, which is comparable to the velocity of the $+$ tertiary. Thus, the relative velocity of the $+$ scattering in this case can be small, down to near zero, leading to a much longer interaction time between the BBH and the tertiary. This makes the $+$ scatterings more prone to cause softening of the BBH rather than hardening. The bottom three panels of the figure show that the ratios $A_{\rm hard,(+),+}/A_{\rm hard,(-),+}$ of 10~au semi-major axis BBHs are smaller than unity. This asymmetry becomes weaker as the BBH and the tertiary get closer to the SMBH, since the orbital velocity about the SMBH gradually cancels out the velocity of the tertiary perturber. The difference in relative velocity (between the BBH and the tertiary) between $+$ and $-$ scatterings becomes smaller. Therefore, the results become more symmetric as the three-body system gets closer to the SMBH.  

\section{Discussion and Conclusions}
Almost all dynamical channels for BBH mergers involve spherical symmetry, such that prograde ($+$) binaries have ($+$),$+$ or  ($+$),$-$ tertiary encounters at the same rate. Likewise the rates of $(-),+$ and $(-),-$ encounters should be comparable.
Thus, most dynamical BBH merger channels (e.g.\ in globular clusters or nuclear star clusters) are expected to have a symmetric projected spin distribution $\chi_{\rm eff}$ around zero \citep[e.g.][]{Carl2018}. AGN disks are one of the few dynamical channels where asymmetries can arise in encounters \citep{qXeff}. If the observed $(q,\chi_{\rm eff})$ anti-correlation in LIGO-Virgo BBH mergers \citep{Tom2021} arises due to AGN disks, then BBHs in AGN disks preferentially  merge biased towards either 1) both $+$ spins \emph{and} ($+$) orbital angular momentum, or 2) both $-$ spins \emph{and} ($-$) orbital angular momentum. 

\begin{figure*}
\includegraphics[width=\linewidth]{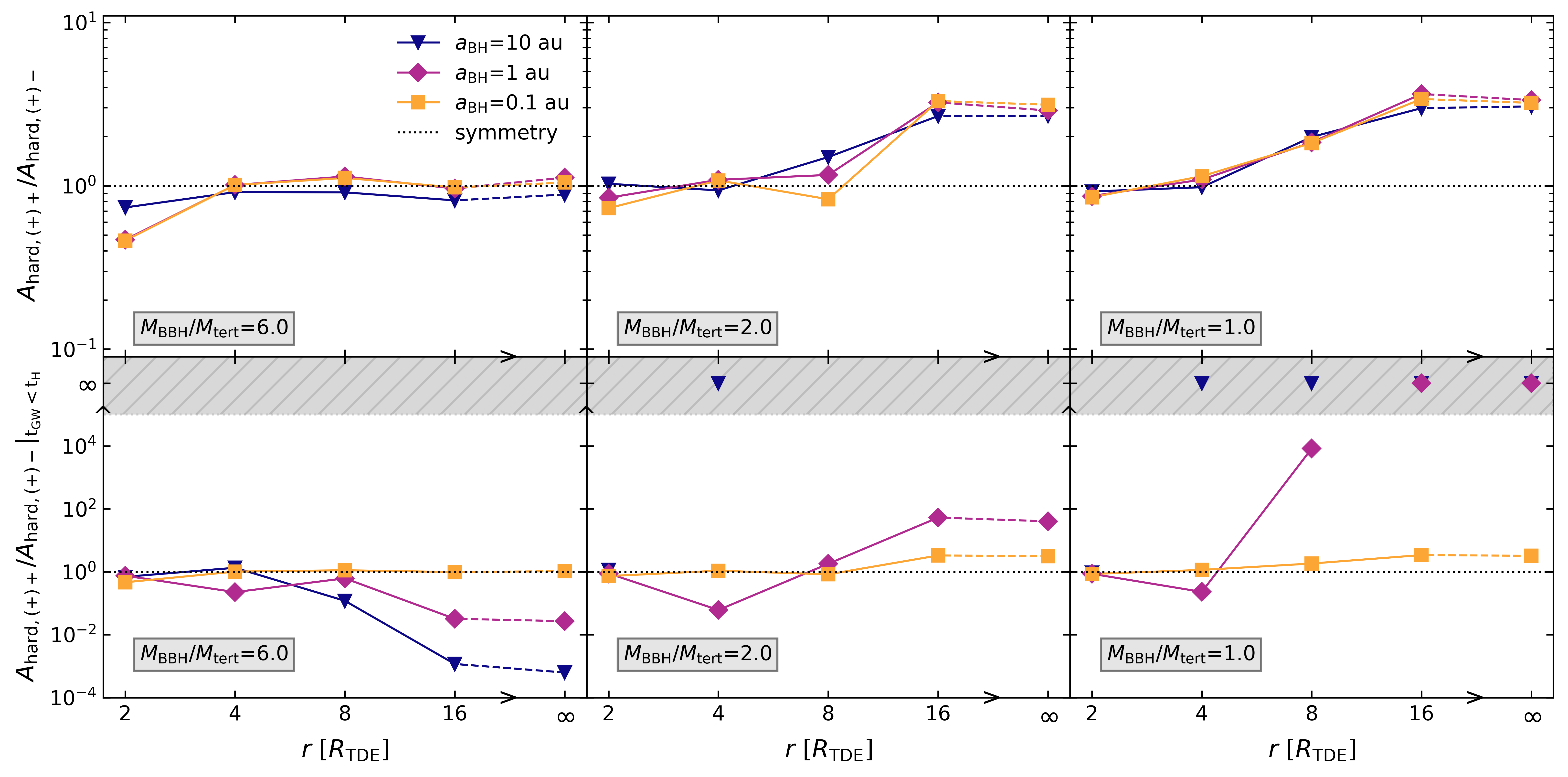}

\caption{Ratio between the phase space areas of hardened prograde and retrograde binaries for the full sample of hardened binaries ({\em top} panels), and for the subset with merger times smaller than the Hubble time ({\em bottom} panels). The three panel columns show the variation with tertiary mass ($M={10}M_\odot$ in the {\em left} panels, $M={30}M_\odot$ in the {\em middle} panels and  $M={60}M_\odot$ in the {\em right} panels). Some points for $a_{\rm BH}=10$~au in the bottom middle and right panels are missing as the corresponding area ratios are not well defined due to both areas being infinitesimally small, while  
    the points at $'\infty'$ on the y-axis (within the shaded area) represent cases where only the area term in the denominator, $(+)-$, is infinitesimally small. 
     Note that the x-axis is logarithmic with a base of 2, but has a discontinuity at the arrow; the points at $'\infty'$ on the x-axis refer to the simulations with no SMBH.
}
\label{fig:ratios}
\end{figure*}

Fig.~\ref{fig:ratios} demonstrates that there is an asymmetry in hardening encounters between ($+,-$) binaries and $+$ tertiary encounters as a function of binary semi-major axis. The asymmetry broadly corresponds to: 1) a preferential hardening of ($+$) binaries at small initial binary separations for similar or larger $+$ tertiary masses, and a preferential hardening of ($-$) binaries at wider separations for smaller $+$ tertiary masses. The asymmetry in ($+,-$) BBH encounters with $+$ tertiaries tends to disappear in the limit of $R_{\rm bin} \rightarrow 2R_{\rm TDE}$. 

If we ascribe the asymmetry in $\chi_{\rm eff}$ observed in LIGO-Virgo mergers \citep{Tom2021,qXeff} to the AGN channel, then the nature of the asymmetry depends on the efficiency of gas hardening and/or the nature of most dynamical encounters. If gas dynamical hardening is efficient, such that $a_{\rm BH}$ is small at the time of the first dynamical encounter, then we expect that more dynamical hardening encounters are between ($+$) binaries and $+$ tertiaries of comparable mass. On the other hand, if gas hardening is not efficient, such that $a_{\rm BH}$ is large at the time of the first dynamical encounter then we expect that most dynamical hardening encounters are due to $(+),-$ or $(-),+$ encounters with lower mass tertiaries. This latter case applies, for example, to a more massive prograde ($+$) binary at a migration trap encountering an inward-migrating less massive tertiary, where in the frame of the prograde binary the encounter sense is $-$, but at low relative velocity ($\sim 50{\rm km/s}$), because of differential disk rotation.

All of our results apply to a single encounter between a prograde BBH ($+$) or retrograde BBH ($-$) with a prograde $+$ tertiary BH. However, we expect multiple encounters in AGN disks are likely in a short time \citep[e.g.][]{Leigh18,Samsing20,Secunda20a,Tagawa20c,McK20b}. Orbital timescales scale as
\begin{equation} t_{\rm orb} \sim 0.5 \mbox{ yr} (M_{\rm SMB H}/10^{8}M_{\odot})(R_{\rm disk}/10^{3}r_{g})^{3/2}, \end{equation} 
so multiple encounters could occur at relatively small separations and relative velocities as modelled above. We have shown that binary dynamical hardening in AGN disks is asymmetric depending on binary initial conditions (binary semi-major axis and direction of binary rotation relative to the tertiary encounter). As a result, the LIGO-Virgo AGN channel should yield an asymmetric distribution of $\chi_{\rm eff}$ for BBH mergers, unlike all other spherically symmetric, dynamical channels (as long as gas hardening is insufficient to bring such binaries into the efficient GW emission regime). For example, a bias towards $\chi_{\rm eff}>0$ mergers could be accounted for by the preferential dynamical hardening of initially wide prograde binaries encountering a less massive migrating tertiary at low relative velocity, either in the bulk disk or at a trap. Such encounters are always in a retrograde sense, but at low relative velocity in the frame of the prograde binary due to differential rotation, and initially wide binaries would imply an inefficient gas hardening process within AGN. A more complete parameter space study will be carried out in future work.

\section{Acknowledgements}
We thank Carl Rodriguez for early discussions of some of the ideas in this paper. YW and RP  acknowledge  support by  NSF  award  AST-2006839. BM \& KESF are supported by NSF AST-1831415 and Simons Foundation Grant 533845. N.W.C.L. gratefully acknowledges support from the Chilean government via Fondecyt Iniciacion Grant 11180005, and acknowledges financial support from Millenium Nucleus $NCN19-058$ (TITANs). M-MML is partly suppported by NSF grant AST18-15461.
\bibliography{refs2}

\begin{thebibliography}{}
\expandafter\ifx\csname natexlab\endcsname\relax\def\natexlab#1{#1}\fi
\providecommand{\url}[1]{\href{#1}{#1}}

\bibitem[{{Bartos} {et~al.}(2017){Bartos}, {Kocsis}, {Haiman}, \&
  {M{\'a}rka}}]{Bartos17}
{Bartos}, I., {Kocsis}, B., {Haiman}, Z., \& {M{\'a}rka}, S. 2017, \apj, 835,
  165

\bibitem[{{Baruteau} {et~al.}(2011){Baruteau}, {Cuadra}, \& {Lin}}]{Baruteau11}
{Baruteau}, C., {Cuadra}, J., \& {Lin}, D.~N.~C. 2011, \apj, 726, 28

\bibitem[{{Bellovary} {et~al.}(2016){Bellovary}, {Mac Low}, {McKernan}, \&
  {Ford}}]{Bellovary16}
{Bellovary}, J.~M., {Mac Low}, M.-M., {McKernan}, B., \& {Ford}, K.~E.~S. 2016,
  \apjl, 819, L17

\bibitem[{{Callister} {et~al.}(2021){Callister}, {Haster}, {Ng}, {Vitale}, \&
  {Farr}}]{Tom2021}
{Callister}, T.~A., {Haster}, C.-J., {Ng}, K. K.~Y., {Vitale}, S., \& {Farr},
  W.~M. 2021, arXiv e-prints, arXiv:2106.00521

\bibitem[{{Dittmann} \& {Miller}(2020)}]{Dittmann2020}
{Dittmann}, A.~J., \& {Miller}, M.~C. 2020, \mnras, 493, 3732

\bibitem[{{Gr{\"o}bner} {et~al.}(2020){Gr{\"o}bner}, {Ishibashi}, {Tiwari},
  {Haney}, \& {Jetzer}}]{Grobner20}
{Gr{\"o}bner}, M., {Ishibashi}, W., {Tiwari}, S., {Haney}, M., \& {Jetzer}, P.
  2020, \aap, 638, A119

\bibitem[{{Leigh} {et~al.}(2014){Leigh}, {Mastrobuono-Battisti}, {Perets}, \&
  {B{\"o}ker}}]{Leigh14}
{Leigh}, N. W.~C., {Mastrobuono-Battisti}, A., {Perets}, H.~B., \& {B{\"o}ker},
  T. 2014, \mnras, 441, 919

\bibitem[{{Leigh} {et~al.}(2018){Leigh}, {Geller}, {McKernan}, {Ford}, {Mac
  Low}, {Bellovary}, {Haiman}, {Lyra}, {Samsing}, {O'Dowd}, {Kocsis}, \&
  {Endlich}}]{Leigh18}
{Leigh}, N.~W.~C., {Geller}, A.~M., {McKernan}, B., {et~al.} 2018, \mnras, 474,
  5672

\bibitem[{{Li} {et~al.}(2021){Li}, {Dempsey}, {Li}, {Li}, \& {Li}}]{Li21}
{Li}, Y.-P., {Dempsey}, A.~M., {Li}, S., {Li}, H., \& {Li}, J. 2021, \apj, 911,
  124

\bibitem[{{Liu} \& {Lai}(2017)}]{Liu2017}
{Liu}, B., \& {Lai}, D. 2017, \apjl, 846, L11

\bibitem[{{McKernan} {et~al.}(2018){McKernan}, {Ford}, {Bellovary},
  {et~al.}}]{McK18}
{McKernan}, B., {Ford}, K.~E.~S., {Bellovary}, J., {et~al.} 2018, \apj, 866, 66

\bibitem[{{McKernan} {et~al.}(2021){McKernan}, {Ford}, {Callister}, {Farr},
  {O'Shaughnessy}, {Smith}, {Thrane}, \& {Vajpeyi}}]{qXeff}
{McKernan}, B., {Ford}, K.~E.~S., {Callister}, T., {et~al.} 2021, arXiv
  e-prints, arXiv:2107.07551

\bibitem[{{McKernan} {et~al.}(2012){McKernan}, {Ford}, {Lyra}, \&
  {Perets}}]{McK12}
{McKernan}, B., {Ford}, K.~E.~S., {Lyra}, W., \& {Perets}, H.~B. 2012, \mnras,
  425, 460

\bibitem[{{McKernan} {et~al.}(2020){McKernan}, {Ford}, \&
  {O'Shaughnessy}}]{McK20b}
{McKernan}, B., {Ford}, K.~E.~S., \& {O'Shaughnessy}, R. 2020, \mnras, 498,
  4088

\bibitem[{{Rodriguez} {et~al.}(2018){Rodriguez}, {Amaro-Seoane}, {Chatterjee},
  \& {Rasio}}]{Carl2018}
{Rodriguez}, C.~L., {Amaro-Seoane}, P., {Chatterjee}, S., \& {Rasio}, F.~A.
  2018, \prl, 120, 151101

\bibitem[{{Samsing} {et~al.}(2020){Samsing}, {Bartos}, {D'Orazio}, {Haiman},
  {Kocsis}, {Leigh}, {Liu}, {Pessah}, \& {Tagawa}}]{Samsing20}
{Samsing}, J., {Bartos}, I., {D'Orazio}, D.~J., {et~al.} 2020, arXiv e-prints,
  arXiv:2010.09765

\bibitem[{{Secunda} {et~al.}(2020){Secunda}, {Bellovary}, {Mac Low}, {Ford},
  {McKernan}, {Leigh}, {Lyra}, {S{\'a}ndor}, \& {Adorno}}]{Secunda20a}
{Secunda}, A., {Bellovary}, J., {Mac Low}, M.-M., {et~al.} 2020, \apj, 903, 133

\bibitem[{{Stone} {et~al.}(2017){Stone}, {Metzger}, \& {Haiman}}]{Stone17}
{Stone}, N.~C., {Metzger}, B.~D., \& {Haiman}, Z. 2017, \mnras, 464, 946

\bibitem[{{Tagawa} {et~al.}(2020{\natexlab{a}}){Tagawa}, {Haiman}, {Bartos}, \&
  {Kocsis}}]{Tagawa20b}
{Tagawa}, H., {Haiman}, Z., {Bartos}, I., \& {Kocsis}, B. 2020{\natexlab{a}},
  \apj, 899, 26

\bibitem[{{Tagawa} {et~al.}(2020{\natexlab{b}}){Tagawa}, {Haiman}, {Bartos}, \&
  {Kocsis}}]{Tagawa20}
---. 2020{\natexlab{b}}, \apj, 899, 26

\bibitem[{{Tagawa} {et~al.}(2020{\natexlab{c}}){Tagawa}, {Haiman}, \&
  {Kocsis}}]{Tagawa20c}
{Tagawa}, H., {Haiman}, Z., \& {Kocsis}, B. 2020{\natexlab{c}}, \apj, 898, 25

\bibitem[{{Tiede} {et~al.}(2020){Tiede}, {Zrake}, {MacFadyen}, \&
  {Haiman}}]{Tiede20}
{Tiede}, C., {Zrake}, J., {MacFadyen}, A., \& {Haiman}, Z. 2020, \apj, 900, 43

\bibitem[{{Wang} {et~al.}(2021){Wang}, {Leigh}, {Liu}, \&
  {Perna}}]{Wang2021SpaceHub}
{Wang}, Y.-H., {Leigh}, N. W.~C., {Liu}, B., \& {Perna}, R. 2021, \mnras, 505,
  1053

\bibitem[{{Yang} {et~al.}(2019){Yang}, {Bartos}, {Gayathri}, {et~al.}}]{Yang19}
{Yang}, Y., {Bartos}, I., {Gayathri}, V., {et~al.} 2019, \prl, 123, 181101

\end{thebibliography}

\end{document}